\documentstyle[epsfig]{article}
\begin{document}
\begin{center}
{\Large \bf The flux-tube model of particle creation in nuclear 
collisions}\,\footnote{Invited talk at Nuclear Matter, Hot and Cold: a Symposium 
in Memory of Judah M. Eisenberg, Tel Aviv, Israel, April 14--16, 1999.
Work supported by the Israel Science Foundation 
under Grant No.~255/96-1 and by the Basic Research Fund of Tel Aviv 
University.}
\vspace{0.3cm}

Benjamin Svetitsky

\vspace{0.3cm}

\noindent
{\it School of Physics and Astronomy, Raymond and Beverly Sackler
Faculty of Exact Sciences, 
Tel Aviv University, 69978 Tel Aviv, Israel}
\vspace{0.5cm}

{\bf Abstract}
\end{center}
\vspace{0.3cm}
\noindent
I review some of the history of the flux-tube model, concentrating on work
done by the Los Alamos--Tel Aviv collaboration on particle creation and
back-reaction in uniform fields and in the central rapidity region.
I discuss the incorporation of more realistic geometry and structure of
the flux tube via application of the dual superconductor model of confinement.
\vspace{1.0cm}

During the decade of my collaboration with Judah, his main area of interest
was the quark--gluon plasma and the efforts to create it in relativistic
heavy-ion collisions.
He was especially drawn to the problem of particle creation in the earliest
stages of a collision, and the process by which these particles might reach
thermodynamic equilibrium.
For a model of particle creation, we focused on the flux-tube model.
It presented us with the practical question of connecting fundamental
quantum field theory with phenomenology, and thus with the theoretical
challenge of connecting quantum field theory with kinetic theory and
hydrodynamics.\footnote{We enjoyed a long-term collaboration 
in this area with Fred Cooper and Emil Mottola of Los Alamos
and with Yuval Kluger, who wrote his Ph.~D.~thesis on the subject at Tel
Aviv and then moved to Los Alamos.}

\section{Kinetic theory of particle creation}
After years of effort and anticipation, the Relativistic Heavy Ion Collider
(RHIC) at Brookhaven will soon begin producing collisions of gold nuclei
in colliding beams, at an energy of 100~GeV per nucleon in each beam.
Theoretical considerations, as well as data from the successful heavy-ion
program at CERN's SPS, lead us to expect that in a central Au--Au collision
the incoming baryons will pass through each other with minimal deflection.
The net outgoing baryon number will thus be concentrated within a couple
of units of rapidity of the beam rapidities, leaving a large central
region to be filled by a baryon-free plasma.
Energy deposition in the central region will take place mainly via soft
mechanisms, although minijet production will play a significant
role.

A QCD-inspired model for soft production in the central region begins
\cite{Biro} with the exchange of soft gluons between the two nuclei as
they pass through each other.
This leaves them charged with color, sources of color electric flux.
The flux fills a tube connecting the receding nuclei, which thus form
a ``color capacitor'' if the field is coherent across the tube, or
a ``color rope'' if it is not.
The electric field strength in either case is proportional to the square
root of the number of gluon exchanges per unit area, and thus $E\sim
A^{1/3}$, as a result of a random walk in color space.

An electric field will pop pairs of charged particles (in this case,
colored particles) out of the vacuum, especially if they are light on
the scale of the available field energy density.
Casher, Neuberger, and Nussinov \cite{Casher}
derived via WKB the rate for particle
creation per unit volume as a function of transverse momentum,
\begin{equation}
\frac{dN}{dt\,dV\,d^2p_\perp}=eE\log\left[1+\exp\left(-
\frac{\pi(m^2+p_\perp^2)}{eE}\right)\right]\ .
\label{CNN}
\end{equation}
(Integrating (\ref{CNN}) over $p_\perp$ gives the famed Schwinger formula
\cite{Schwinger}.)

If the flux tube is narrow to begin with, as in the case of $e^+e^-$
annihilation into quarks or of $pp$ scattering, the quarks produced via
(\ref{CNN}) will break the tube and begin an inside--outside cascade.
For nucleus--nucleus collisions, however, the geometry more closely resembles
that of the idealized problem of a homogeneous electric field created by
infinite, parallel capacitor plates.
Then the created particles form a uniform current that screens the field
gradually according to (Abelian) Maxwell's equations,
\begin{equation}
\frac{dE}{dt}=-j\ .
\label{Maxwell}
\end{equation}
This is called {\em back-reaction.}
A straightforward approach to the particle dynamics
is given by kinetic theory, where the pair
creation rate (\ref{CNN}) appears as a source term for the Vlasov
equation \cite{Kajantie,Bialas},
\begin{equation}
\frac{\partial f}{\partial t}+eE\frac{\partial f}{\partial p_z}=
eE\log\left[1+\exp\left(-
\frac{\pi(m^2+p_\perp^2)}{eE}\right)\right]\delta(p_z)\ .
\label{Vlasov}
\end{equation}
(The particle density $f$ feeds back into the Maxwell equation (\ref{Maxwell})
through the current.)
This gives a beautiful description of matter creation and subsequent
plasma oscillations.
The drawback of this approach, of course, is that the use of the classical
Vlasov formalism lacks fundamental justification.

\section{Enter quantum field theory}
Quantum field theory allows a first-principles approach to the problem
\cite{CM,bosons}.
One begins by taking the number of flavors $N_f$ to infinity in order to justify
a classical, mean-field approximation for $E$.
If we begin with a scalar field in one dimension,
the ingredients of the theory are then the Klein--Gordon equation
for the matter field,
\begin{equation}
[(\partial ^\mu +ieA^\mu )(\partial _\mu +ieA_\mu )+ m^2]\Phi (x)=0\ ,
\label{KG}
\end{equation}
and the semiclassical Maxwell equation,
\def\eval#1{\left\langle#1\right\rangle}
\begin{equation}
\partial_\mu F^{\mu \nu}=\eval{0| j^\nu | 0}\ .
\label{MaxwellSC}
\end{equation}
A Fourier expansion of $\Phi$,
\begin{equation}
\Phi (x,t) = \int{dk\over2\pi}\,\left[f_k(t)a_k+
f_k^{\ast} (t) b_{-k}^{\dagger}\right]e^{i{k}{x}}\ ,
\label{Fourier}
\end{equation}
inserted into (\ref{KG}), gives evolution equations for the
Fourier amplitudes $f_k$,
\begin{equation}
{d^2 f_k(t)\over dt^2}+\omega _k^2 (t)f_k(t)=0\ ,
\label{f_eqn}
\end{equation}
where
\begin{equation}
\omega _k (t)^2 \equiv [k-eA(t)]^2 +m^2\ .
\end{equation}
The canonical commutation relations for $\Phi$ imply that $f_k$ can be written as
\begin{equation}
f_{{k}} (t)={1\over\sqrt {2 \Omega_{{k}}(t)}}\,e^{-i\int ^{t} \Omega_{{k}}(t')dt'}\ .
\end{equation}
Inserting this into (\ref{f_eqn}) gives equations for
the effective frequencies $\Omega_{{k}}(t)$,
\begin{equation}
\Omega^{2}_{k}(t) =-\frac{\ddot{\Omega}_{k}}{2\Omega_{k}}
+{3\over4}\left(\frac{\dot{\Omega}_{k}}{\Omega_k}\right)^2 
+ {\omega}^{2}_{k}(t)\ .
\label{Omega}
\end{equation}
We can write the current $j$ in terms of the amplitudes $f_k$, and
(\ref{MaxwellSC}) becomes the subtracted, renormalized equation
\begin{equation}
\ddot{A}=e\int{dk\over2\pi}(k-eA)
\left[{1\over\Omega_k(t)}-{1\over\omega _k(t)}\right]\ .
\label{Maxwell2}
\end{equation}
Equations (\ref{Omega}) and (\ref{Maxwell2}) can be integrated numerically.

\begin{figure}[htb]
\vskip 5mm
\centerline{\epsfig{file=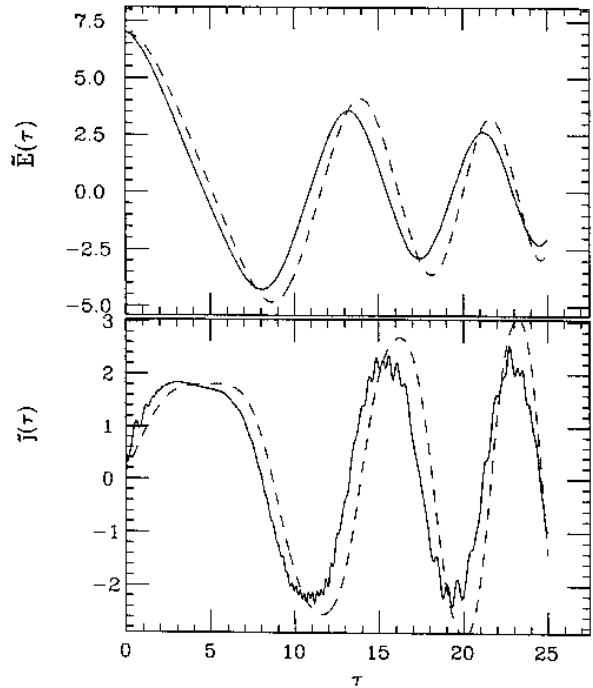,width=8cm}}
\caption{Plasma oscillations of bosons in three dimensions}
\label{fig1}
\end{figure}

We applied this formalism to the creation of bosons \cite{bosons} and
fermions \cite{fermions} in one and three \cite{ijmpe} dimensions.
For illustration, I show in fig.~\ref{fig1} the time evolution (in scaled
variables) of a system of bosons in three dimensions.
Particles created at early times are accelerated into plasma oscillations,
with continued particle creation whenever the field is non-zero.
A comparison to the results of 
kinetic theory [a slight modification of (\ref{Vlasov}): dashed
curves] shows that
the latter approximates the field-theoretic solution well.
Space limitations prevent me from showing the beautiful results of comparing
the phase space distributions of produced particles in the two approaches.
\section{Extensions}
The idealized problem of infinite, static capacitor plates was of course only
a first step in development of the flux tube model.
With an eye towards nucleus--nucleus collisions, we extended the analysis
\cite{boost}
to the problem of capacitor plates separating at the speed of light,
$z=\pm ct$.
As is natural in the context of the central rapidity region \cite{CFS},
we assumed that the initial conditions are invariant under longitudinal
boosts so that the time development depnds only on the invariant, comoving
time $\tau=\sqrt{t^2-z^2}$.
The phase-space distribution of the produced particles now gives the rapidity
distribution $dN/dy$, and analysis of the energy-momentum tensor gives
the energy density and equivalent temperature as functions of $\tau$ (though
there is as yet no mechanism for equilibration).
As it turns out, the results are again well approximated by kinetic theory.

The geometry to this point is still that of an idealized
system of infinite transverse extent.
Upon solving the problem in a cylinder of finite radius (a computationally
formidable problem due to nonlinear mode mixing), Judah found \cite{JMEcyl}
that the agreement between field theory and kinetic theory is at best 
qualitative.

As mentioned, the semiclassical approximation presented above finds its
justification in the limit of large $N_f$.
The next order in $1/N_f$ introduces particle--particle collisions and thus
a mechanism for relaxation to thermal equilibrium \cite{largeN}.
Unfortunately, the resulting integro-differential equations are (so far)
computationally unmanageable.
Kluger, Mottola, and Eisenberg \cite{Vlasov} later eschewed this approach and 
showed, in what was to be Judah's last paper,  that equilibration can arise even
in the lowest-order mean field theory shown above if one allows for {\em
dephasing} \cite{Habib} of the highest-frequency quantum oscillations.
The connection between reversible quantum mechanics and irreversible kinetic
theory has long been an area of mystery and of hard work.
In this last paper, a Vlasov equation non-local in time is actually
derived for the pair-creation problem.\footnote{The paper is a mathematical {\em tour de force\/} of
Airy functions, uniform asymptotic expansions, etc.
Anyone who knew Judah's great love for mathematical physics of the Morse \&
Feshbach variety will recognize Judah walking in its pages.}

\section{The flux tube of the dual superconductor}

Judah, as noted, applied the field-theoretic formalism to particle production
in a cylindrical flux tube of finite, fixed radius.
The QCD flux tube, however, is a dynamical object, governed by field equations.
't~Hooft and Mandelstam noted long ago that flux tubes arise naturally in
superconductors via the Meissner effect; just as magnetic monopoles would
be confined by a magnetic flux tube in a superconductor, quarks with their
{\em electric\/} color charge would be confined by an electric flux tube
if the QCD vacuum has the structure of a {\em dual\/} superconductor.
Melissa Lampert and I have taken the first step of studying the dynamics
of classical charges moving in an electric flux tube and the reaction
of the flux tube in this model \cite{LS}.

To specify the dual superconductor model, we begin with
Maxwell's equations coupled to both magnetic and electric currents,
\begin{eqnarray}
   \partial_\mu F^{\mu\nu} &=& j_e^\nu \label{egauss4}\\
   \partial_\mu \tilde F^{\mu\nu} &=& j_g^\nu \label{mgauss4}\ .
\end{eqnarray}
Eq.~(\ref{mgauss4}) is no longer just a Bianchi identity; thus
a vector potential can be introduced only if a new term is
added to take care of the magnetic current,
\begin{equation}
F^{\mu\nu} = \partial^\mu A^\nu - \partial^\nu A^\mu 
      + \epsilon^{\mu\nu\lambda\sigma} G_{\lambda\sigma}\ ,
\end{equation}
\begin{equation}
   G^{\mu\nu} = - n^\mu (n \cdot \partial)^{-1} j_g^\nu\ .
\end{equation}
This vector potential can be coupled to electric charges as usual;
in order to introduce {\em magnetic\/} charges, one introduces a
dual potential \cite{Zwanziger} via
\begin{equation}
   \tilde F^{\mu\nu} = \partial^\mu B^\nu - \partial^\nu B^\mu  
      + \epsilon^{\mu\nu\lambda\sigma} M_{\lambda\sigma}\ ,
\end{equation}
\begin{equation}
   M^{\mu\nu} = - n^\mu (n \cdot \partial)^{-1} j_e^\nu\ .
\end{equation}
Now we can write a model for the monopoles, for which the simplest is an
Abelian Higgs theory \cite{Suzuki,Osaka},
\begin{equation}
   D_\mu^B D^{\mu B}\psi + \lambda(|\psi|^2-v^2)\psi = 0 \label{Higgs}\ ,
\end{equation}
where
\begin{equation}
   D^B_\mu \equiv \partial_\mu - igB_\mu\ .
\end{equation}
This theory should produce the desired magnetic condensate to confine electric
charge.
The magnetic current appearing in (\ref{mgauss4}) is
\begin{equation}
j_g^{\mu}=2g\,{\rm Im}\,\psi^*D^{\mu B}\psi\ .
\end{equation}
For the electric charges, we take simple two-fluid MHD (see \cite{LS} for 
details).

\begin{figure}[htb]
   \begin{center} 
\vskip -5mm
   \mbox{\epsfig{file=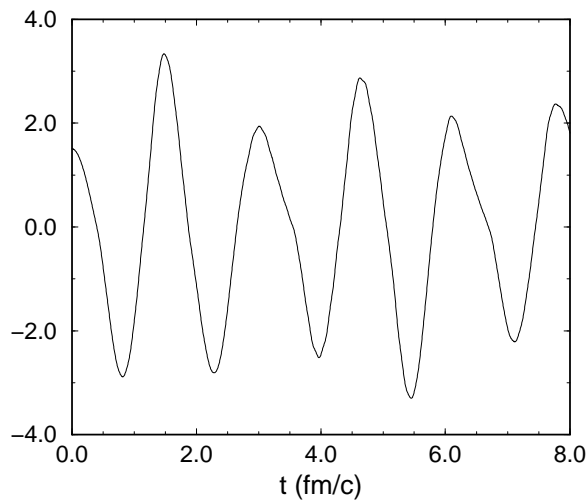,width=3.5in}}
   \end{center}
\vskip -5mm
\caption{Plasma oscillations in the Type~I dual superconductor with
    $\omega_p < m_V$: the electric field on the axis of the flux tube.}
\label{fig2}
\end{figure}

The parameters of the theory may be adjusted to put the superconductor into
the Type~I or the Type~II regime; the density of the charged fluid will make
the plasma frequency $\omega_p$ larger or smaller than the mass $m_V$ of the
gauge boson in this Higgs theory.
I show in fig.~\ref{fig2} plasma oscillations in the flux tube for a Type~I
superconductor with $\omega_p<m_V$.
The oscillations are made nonlinear by the reaction of the flux tube, which 
tries to close up when the field is weak.

This is a first step towards the inclusion of the dynamics of the flux tube
in a model of particle creation.
The latter might be included, as discussed above, as a kinetic theory based
on a Vlasov equation or via a full field-theoretic treatment.
As argued in \cite{LS}, however, the simplest Higgs theory (\ref{Higgs})
doesn't contain enough {\em dynamical}
confinement physics to produce good phenomenology.

\end{document}